\documentstyle[12pt]{article}

\def\beq{\begin{equation}} \def\eeq{\end{equation}}
\def\bea{\begin{eqnarray}} \def\eea{\end{eqnarray}}
\let\nn=\nonumber
\def\beann{\begin{eqnarray*}} \def\eeann{\end{eqnarray*}}

\let\a=\alpha \let\be=\beta \let\g=\gamma \let\de=\delta
\let\e=\varepsilon   
  \let\la=\lambda \let\m=\mu
  \let\p=\pi  \let\s=\sigma
 
\let\ph=\varphi \let\Ph=\phi  
 \let\Si=\Sigma 
\let\La=\Lambda  \let\D=\Delta

\let\qd=\quad \let\qqd=\qquad \def\qqqd{\qquad\qquad}

\def\tst#1{{\textstyle #1}}
\def\dst#1{{\displaystyle #1}}
\def\sst#1{{\scriptstyle #1}}

\def\0{\over } \def\1{\vec }     \def\2{{1\over2}} \def\4{{1\over4}}
\def\5{\bar }  \def\6{\partial } \def\7#1{{#1}\llap{/}}

\def\<{\langle } \def\>{\rangle }

\let\auf=\uparrow \let\ab=\downarrow

\def\i{{\rm i}}

  \def\CL{{\cal L}}
 \def\CT{{\cal T}} \def\CO{{\cal O}}
\def\tr{\mbox{tr}} 
\def\str{\mbox{str}}

\def\sign{\mbox{sign}} \def\End{\mbox{End}} \def\id{\mbox{id}}
\def\mod{\mbox{\,mod\,}}

\def\smod{\mbox{\scriptsize mod}}

\newfont{\fettohne}{cmssbx10 scaled 1000}
\newfont{\gfettohne}{cmssbx10 scaled 1200}
\newfont{\kfettohne}{cmssbx10 scaled 700}

\def\MC{\mbox{{\gfettohne C}}}

\begin{document}
\thispagestyle{empty}
\begin{center}
{\Large {\bf Fermionic representations of integrable lattice systems\\}}
\vspace{7mm}
{\large Frank G\"{o}hmann$^\dagger$\footnote
{e-mail: goehmann@insti.physics.sunysb.edu}
and Shuichi Murakami$^\ddagger$\footnote[4]
{e-mail: murakami@appi.t.u-tokyo.ac.jp}}\\
\vspace{5mm}
$^\dagger$Institute for Theoretical Physics,\\ State University of New
York at Stony Brook,\\ Stony Brook, NY 11794-3840, USA\\
$^\ddagger$Department of Applied Physics, University of Tokyo,\\
Hongo 7-3-1, Bunkyo-ku, Tokyo 113, Japan
\vspace{7mm}

{\large {\bf Abstract}}
\end{center}
\begin{list}{}{\addtolength{\rightmargin}{10mm}
               \addtolength{\topsep}{-5mm}}
\item
We develop a general scheme for the use of Fermi operators within
the framework of integrable systems. This enables us to read off
a fermionic Hamiltonian from a given solution of the Yang-Baxter
equation and to express the corresponding $L$-matrix and the generators
of symmetries in terms of Fermi operators. We illustrate our approach
through a number of examples. Our main example is the algebraic Bethe
ansatz solution of the Hubbard model in the infinite coupling limit.
\end{list}

\clearpage

\subsection*{Introduction}
The purpose of this article is to consider graded vector
spaces and the graded Yang-Baxter algebra in such a way that the
explicit construction of integrable models in terms of
Fermi operators becomes easy. We will present simple general formulae,
which will enable expression of Hamiltonian, $L$-matrix, Yang-Baxter
algebra and generators of symmetries in terms of Fermi operators, once
a solution of the Yang-Baxter equation is given. 

The material developed below has its
origin in two articles of Kulish and Sklyanin \cite{KuSk82,Kulish85},
where fundamental, graded integrable lattice systems were considered
for the first time. Although most of the basic ideas are outlined in
\cite{KuSk82,Kulish85}, these articles remain rather sketchy as far
as concrete representations in terms of Fermi operators are concerned.
We shall try to explain in the following that this is a topic of its
own interest.

We would like to emphasize that the construction of fermionic
representations is a subject which has to be seen separately from the
construction of integrable models invariant under Lie superalgebras
or their deformations (cf e.g.\ \cite{BGZD94,DGLZ95,Maassarani95}).
Although Lie superalgebra invariant models are most naturally
represented in terms of Fermi operators, there is no need to do so.
On the other hand, there are models, such as the Hubbard model, or
its infinite coupling limit, to be treated below, which are not
invariant under graded algebras, but have their most interesting
interpretation in terms of Fermi operators. The possibility of
connecting a given solution of the Yang-Baxter equation with a
fermionic representation is only restricted by a rather weak
compatibility condition (eq.\ (\ref{comp})). Thus there may be
different fermionic representations of the same model, which correspond
to different gradings.

One possibility to connect spin models with fermionic models, which
is frequently encountered in the literature, is the Jordan-Wigner
transformation. This transformation is mostly applied on the level of
Hamiltonians. Applying it to the Yang-Baxter algebra is a feasible, yet
cumbersome task. The Jordan-Wigner transformation led to important
progress, e.g.\ in the understanding of the Hubbard model
\cite{Shastry88b,OWA87}. We shall argue however that it is an
unnecessary element for the construction of fermionic representations
and that the route taken in the present article leads to a clearer
understanding of the issue. The Jordan-Wigner transformation does
not preserve the boundary conditions. It may obstruct symmetries, and
is not easily generalizable to an arbitrary number of internal degrees
of freedom.

The plan of the present paper is as follows. We first present some
necessary mathematical preliminaries. We introduce the parity, a
concept of odd and even on the {\it basis} of a finite dimensional
vector space, which is then called a graded vector space. Later this
will allows us to make contact with Fermi operators. We extend the
concept of parity to endomorphisms of the vector space and to tensor
powers of endomorphisms. The central definition is equation
(\ref{defej}), which provides an embedding of a ``local'' basis
$\{e_\a^\be\}$ of endomorphisms into a ``chain of $L$ sites'',
$\{e_\a^\be\} \rightarrow \{{e_j}_\a^\be\}$, in such a way that the
${e_j}_\a^\be$, depending on the value of $\a$ and $\be$, either
commute or anticommute. The ${e_j}_\a^\be$ are ``graded analogs of spin
operators''.  It will turn out later that they can be directly replaced
by fermionic projection operators. This is an advantage of considering
the grading on a basis rather than on coordinates. Before turning to
fermions, we have to develop the quantum inverse scattering method
using the basis $\{{e_j}_\a^\be\}$. We try to keep the analogy to the
non-graded case as close as possible and try to emphasize in our
presentation the modifications due to the grading. We discuss the
general case of global symmetries, which, due to the grading are
subalgebras of Lie superalgebras. Later, after having introduced Fermi
operators, we look at the example of gauge transformations, where we
also discuss the local case.

Finally, we illustrate our approach with a number of examples. Most
of these examples will be familiar to the reader. They were chosen
for this reason. The last example, however, is new and of its own
interest. Our approach allows us to identify the $R$-matrix of the
Hubbard model in the infinite coupling limit as a member of a family
of $R$-matrices of su(N) XX chains, which was recently proposed by
Maassarani and Mathieu \cite{MaMa97a}. Since the example is new, we work
it out in some detail. We perform an algebraic Bethe ansatz for the
model and investigate its symmetries and the question of completeness
of the algebraic Bethe ansatz eigenstates.
\subsection*{Graded vector spaces}
Within the formalism of second quantization a physical lattice system
can be entirely described by a set of creation and annihilation
operators $a_{j \a}^+$, $a_{j \a}$ of particles at site $j$ of the
lattice. Depending on the index $\a$ these particles may be bosons
or fermions. Accordingly, they commute or anticommute at different
sites,
\beq \label{bofe}
     a_{j \a} a_{k \be} \pm a_{k \be} a_{j \a} = 0 \qd, \qd j \ne k \qd.
\eeq
Let us consider systems with a finite number of states per site, say
$m$ bosonic and $n$ fermionic states, $\a = 1, \dots, m + n$.
We can look for matrix representations of the operators $a_{j \a}$
acting on tensor products of the local space $\MC^{m + n}$. For spinless
fermions, for instance, such a representation is provided by the
Jordan-Wigner transformation. Unoccupied sites are bosonic, occupied
sites are fermionic.

If we want to describe the general situation, it turns out to be useful
to reverse the above reasoning. Let us start from a local space of
states $V$, which is then isomorphic to $\MC^{m + n}$, and let us impose
an additional structure, the parity, from the outset. Let $V = V_0
\oplus V_1$, and call $v_0 \in V_0$ even, $v_1 \in V_1$ odd. The
subspaces $V_0$ and $V_1$ are called the homogeneous components of
$V$. The parity $p$ is a function $V_i \longrightarrow Z_2$ defined
on the homogeneous components of $V$,
\beq
     p(v_i) = i \qd, \qd i = 0, 1 \qd, \qd v_i \in V_i \qd.
\eeq
The vector space $V$ equipped with this structure is called a graded
vector space or superspace. Fix a basis $\{e_1, \dots, e_{m + n}\}$
of definite parity. Let $p(\a) := p(e_\a)$. Since we want to construct
an algebra of commuting and anticommuting {\it operators}, we have to
extend the concept of parity to operators in End($V$) and to tensor
products of these operators. Let $e_\a^\be \in \End (V)$, $e_\a^\be
e_\g = \de_\g^\be e_\a$. $\{e_\a^\be\}$ is a basis of End($V$). If we
represent $e_\g$ as a column matrix with only non-vanishing entry 1 in
row $\g$, then $e_\a^\be$ is an $(m + n) \times (m + n)$ matrix with
only non-vanishing entry 1 in row $\a$ and column $\be$. Define the
parity of rows and columns,
\beq
     p_r (e_\a^\be) = p(\a) \qd, \qd p_c (e_\a^\be) = p(\be)
\eeq
and the parity of matrices
\beq
     p(e_\a^\be) = p(\a) + p(\be) \qd.
\eeq
With this definition End($V$) becomes a graded vector space.

Take $X$, $Y$ from the homogeneous components of End($V$), and define
the superbracket
\beq
     [X,Y]_\pm = XY - (-1)^{p(X)p(Y)} YX \qd.
\eeq
Extend the superbracket linearly in both of its arguments to
End($V$). Then, End($V$) endowed with the superbracket becomes the
Lie superalgebra \linebreak gl($m|n$). Note that the above definition
of a superbracket makes sense in any graded algebra. We will also use
it in the context of graded tensor powers of End($V$), which is our
next issue.

The notion of a grading may be extended to the $L$-fold tensor product
of End($V$), setting
\bea
     p_r (e_{\a_1}^{\be_1} \otimes \dots \otimes e_{\a_L}^{\be_L})
          & = &  p(\a_1) + \dots + p(\a_L) \qd,\\
     p_c (e_{\a_1}^{\be_1} \otimes \dots \otimes e_{\a_L}^{\be_L})
          & = &  p(\be_1) + \dots + p(\be_L) \qd,\\
     p (e_{\a_1}^{\be_1} \otimes \dots \otimes e_{\a_L}^{\be_L})
          & = &  p(\a_1) + p(\be_1) + \dots + p(\a_L) + p(\be_L) \qd.
\eea
It can be seen from the last formula, that homogeneous elements
$A = A^{\a_1 \dots \a_L}_{\be_1 \dots \be_L} e_{\a_1}^{\be_1} \otimes
\dots \otimes e_{\a_L}^{\be_L}$ of $(\End (V))^{\otimes L}$ with
parity $p(A)$ are characterized by the equation
\beq
     (-1)^{\sum_{j=1}^L (p(\a_j) + p(\be_j))}
        A^{\a_1 \dots \a_L}_{\be_1 \dots \be_L} =
     (-1)^{p(A)} A^{\a_1 \dots \a_L}_{\be_1 \dots \be_L} \qd.
\eeq
This implies that $AB$ is homogeneous with parity
\beq \label{homab}
     p(AB) = p(A) + p(B) \qd,
\eeq
if $A$ and $B$ are homogeneous.

The above definitions allow us to introduce another tensor product
structure $\otimes_s$, which is called the graded or supertensor
product, on exterior powers of End($V$). Choose $v$ and $w$ from the
homogeneous components of $(\End (V))^{\otimes k}$ and
$(\End (V))^{\otimes l}$, respectively. Then, by definition,
\beq \label{deftens}
     v \otimes_s w = (-1)^{p(v)p_r (w)} \, v \otimes w \qd.
\eeq
As a simple consequence of this definition, the supertensor product
is associative, $(u \otimes_s v) \otimes_s w  = u \otimes_s
(v \otimes_s w)$. This follows first for $u$, $v$, $w$ taken from
the homogeneous components of certain powers of End($V$), and then
by linearity for arbitrary $u$, $v$ and $w$.

Here comes the central definition of the paper. The supertensor
product induces an embedding of $e_\a^\be$ into
$(\End (V))^{\otimes L}$, \bea \label{defej}
     {e_j}_\a^\be & = & I_{m + n}^{\otimes_s (j - 1)} \otimes_s
           e_\a^\be \otimes_s I_{m + n}^{\otimes_s (L - j)} \\
	   \label{defej2}
	   & = & (-1)^{(p(\a) + p(\be)) \sum_{k = j + 1}^L p(\g_k)}
           I_{m + n}^{\otimes (j - 1)} \otimes
           e_\a^\be \otimes e_{\g_{j+1}}^{\g_{j+1}} \otimes \dots
	   \otimes e_{\g_L}^{\g_L} \qd.
\eea
$I_{m + n}$ in this equation denotes the $(m + n) \times (m + n)$
unit matrix.  In the second equation summation over double tensor
indices is understood. The index $j$ on the left hand side of
(\ref{defej}) will be called site index. The matrices ${e_j}_\a^\be$
realize relations of the form (\ref{bofe}). For $j \ne k$ we find
\beq \label{coantico}
     {e_j}_\a^\be {e_k}_\g^\de = (-1)^{(p(\a) + p(\be))(p(\g) + p(\de))}
          {e_k}_\g^\de {e_j}_\a^\be \qd.
\eeq
It follows from eq.\ (\ref{defej2}) that ${e_j}_\a^\be$ is homogeneous,
and that
\beq
     p({e_j}_\a^\be) = p(\a) + p(\be) \qd.
\eeq
Hence, in agreement with intuition, eq.\ (\ref{coantico}) says that
odd matrices mutually anticommute, whereas even matrices commute with
each other as well as with the odd matrices. For products of matrices
${e_j}_\a^\be$ which are acting on the same site (\ref{defej2}) implies
\beq \label{samesite}
     {e_j}_\a^\be {e_j}_\g^\de = \de_\g^\be {e_j}_\a^\de \qd.
\eeq
Using the superbracket, (\ref{coantico}) and (\ref{samesite}) may
be combined to
\beq
     [{e_j}_\a^\be,{e_k}_\g^\de]_\pm =
	  \de_{jk} \left( \de_\g^\be {e_j}_\a^\de -
	  (-1)^{(p(\a) + p(\be))(p(\g) + p(\de))}
          \de_\a^\de {e_j}_\g^\be \right) \qd.
\eeq 
The right hand side of the latter equation with $j = k$ gives the
structure constants of the Lie superalgebra gl($m|n$) with respect
to the basis $\{{e_j}_\a^\be\}$.
\subsection*{The permutation operator}
The permutation operator plays an important role in the construction of 
local integrable lattice models. In the graded case it requires the
following modifications of signs,
\beq \label{defp}
     P_{jk} = (-1)^{p(\be)} {e_j}_\a^\be {e_k}_\be^\a \qd.
\eeq
This operator induces the action of the symmetric group $S^L$ on
the site indices of the matrices ${e_j}_\a^\be$. The following
properties are easily verified, they follow from (\ref{coantico})
and (\ref{samesite}),
\beq
     \begin{array}{r@{\qqqd}r@{\qd = \qd}l}
     (a) & P_{kj} & P_{jk} \qd, \\
     (b) & P_{jj} & (m - n) \id \qd, \\
     (c) & P_{jk}^2 & \id \qd, \qd j \ne k \qd, \\
     (d) & P_{jk} {e_k}_\a^\be & {e_j}_\a^\be P_{jk} \qd, \\
     (e) & P_{jk} {e_l}_\a^\be & {e_l}_\a^\be P_{jk} \qd,
           \qd j \ne l \ne k \qd.
     \end{array}
\eeq
Because of ($d$) and ($e$) $P_{jk}$ generates a faithful representation
of $S^L$,
\beq
     P_{jk} P_{kl} = P_{jl} P_{jk} = P_{kl} P_{jl} \qd.
\eeq
Let $L = 2$. Then
\beq
     P_{12} = (-1)^{p(\be)} {e_1}_\a^\be {e_2}_\be^\a
            = (-1)^{p(\a)p(\be)} e_\a^\be \otimes e_\be^\a
	    = (-1)^{p(\a)p(\be)} \de_\de^\a \de_\g^\be
	      e_\a^\g \otimes e_\be^\de \qd.
\eeq
From the right-hand side of this equation we can read off the matrix
elements of $P_{12}$ with respect to the canonical basis of
End($V \otimes V$).
\subsection*{The graded Yang-Baxter algebra}
\unitlength .85mm
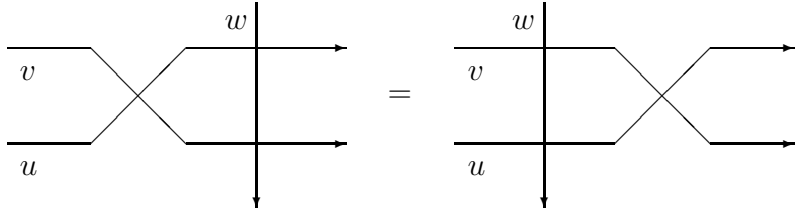
\begin{figure}

\begin{picture}(140,50)


\put(12,15){\line(1,0){13}}
\put(12,30){\line(1,0){13}}
\put(25,15){\line(1,1){15}}
\put(25,30){\line(1,-1){15}}
\put(40,15){\vector(1,0){25}}
\put(40,30){\vector(1,0){25}}
\put(51,37){\vector(0,-1){32}}

\put(14,10){$u$}
\put(14,25){$v$}
\put(46,33){$w$}

\put(71.5,21.5){=}

\put(122,15){\vector(1,0){13}}
\put(122,30){\vector(1,0){13}}
\put(107,15){\line(1,1){15}}
\put(107,30){\line(1,-1){15}}
\put(107,15){\line(-1,0){25}}
\put(107,30){\line(-1,0){25}}
\put(96,37){\vector(0,-1){32}}

\put(84,10){$u$}
\put(84,25){$v$}
\put(91,33){$w$}

\end{picture}
\caption{The Yang-Baxter equation is most easily memorized in
graphical form}
\end{figure}
In the present context it is most suitable to interpret the Yang-Baxter
equation as a set of functional equations for the matrix elements
of an $(m + n)^2 \times (m + n)^2$-matrix $R(u,v)$. We may represent it
in graphical form as shown in Figure 1, where each vertex corresponds
to a factor in the equation
\beq \label{ybe}
     R_{\a' \be'}^{\a \be} (u,v) R_{\a'' \g'}^{\a' \g} (u,w)
     R_{\be'' \g''}^{\be' \g'} (v,w) =
     R_{\be' \g'}^{\be \g} (v,w) R_{\a' \g''}^{\a \g'} (u,w)
     R_{\a'' \be''}^{\a' \be'} (u,v) \qd.
\eeq
Note that there is a direction assigned to every line in figure 1,
which is indicated by the tips of the arrows. Therefore every vertex
has an orientation, and vertices and $R$-matrices can be identified
according to figure 2, where indices have been supplied to a vertex.
Summation is over all inner lines in figure 1.
\begin{figure}

\begin{picture}(140,40)

\put(25,20){$R^{\a \be}_{\g \de} (u,v) \qd =$}

\put(70,21.5){\vector(1,0){26}}
\put(82,33.5){\vector(0,-1){26}}

\put(71,17){$u$}
\put(78,30.5){$v$}

\put(64,21){$\sst{\a}$}
\put(81,38){$\sst{\be}$}
\put(81,1){$\sst{\de}$}
\put(100,21){$\sst{\g}$}

\end{picture}

\caption{Identification of the $R$-matrix with a vertex}
\end{figure}

Starting from the Yang-Baxter equation we will construct a fundamental
graded representation of the Yang-Baxter algebra. For comparison
let us briefly recall the non-graded case ($n = 0$). Define
$\check R (u,v)$ as
\beq
     \check R^{\a \be}_{\g \de} (u,v) = R^{\be \a}_{\g \de} (u,v) \qd.
\eeq
Introduce the $L$-matrix at site $j$,
\beq
     {L_j}^\a_\be (u,v) = R^{\a \g}_{\be \de} (u,v) {e_j}_\g^\de \qd.
\eeq
Then multiplication of the Yang-Baxter equation (\ref{ybe}) by
${e_j}_\g^{\g''}$ implies that
\beq \label{yba}
     \check R(u,v) \left( L_j (u,w) \otimes L_j (v,w) \right) =
     \left( L_j (v,w) \otimes L_j (u,w) \right) \check R(u,v) \qd,
\eeq
where the tensor product is now a tensor product between matrices,
according to the convention $(A \otimes B)^{\a \g}_{\be \de} =
A^\a_\be B^\g_\de$. We may replace $L_j$ in (\ref{yba}) by some
matrix $T$ and may interpret it as defining an abstract algebra for
the matrix elements of $T$. This algebra is called a Yang-Baxter
algebra with $R$-matrix $R$. $L_j$ is called its fundamental
representation. Since $L$-matrices at different sites commute, any
product of $L$-matrices with different site indices is another
representation of the same Yang-Baxter algebra.

The construction of a graded Yang-Baxter algebra and its fundamental
representation requires only minimal modifications of the above scheme.
Let us assume we are given a solution of (\ref{ybe}), which is
compatible with the grading in the sense that
\beq \label{comp}
     R_{\g \de}^{\a \be} (u,v) = (-1)^{p(\a) + p(\be) + p(\g) + p(\de)}
     R_{\g \de}^{\a \be} (u,v) \qd.
\eeq
Define a graded $L$-matrix  at site $j$ as
\beq \label{defgl}
     {\CL_j}^\a_\be (u,v) = (-1)^{p(\a) p(\g)}
         R^{\a \g}_{\be \de} (u,v) {e_j}_\g^\de \qd.
\eeq
Eq.\ (\ref{comp}) implies that the matrix elements of ${\CL_j} (u,v)$
are of definite parity,
\beq
     p( {\CL_j}^\a_\be (u,v)) = p(\a) + p(\be) \qd,
\eeq
and that they commute as
\beq \label{ljlk}
     {\CL_j}^\a_\be (u,v) {\CL_k}^\g_\de (w,z) =
          (-1)^{(p(\a) + p(\be))(p(\g) + p(\de))}
	  {\CL_k}^\g_\de (w,z) {\CL_j}^\a_\be (u,v) \qd.
\eeq
It further follows from the Yang-Baxter equation (\ref{ybe}) and
from eq.\ (\ref{comp}) that
\beq \label{gyba}
     \check R(u,v) \left( \CL_j (u,w) \otimes_s \CL_j (v,w) \right) =
     \left( \CL_j (v,w) \otimes_s \CL_j (u,w) \right) \check R(u,v) \qd.
\eeq
In analogy to the non-graded case above, the supertensor product
in this equation is to be understood as a supertensor product
of matrices with non-commuting entries,
$(A \otimes_s B)^{\a \g}_{\be \de} = (-1)^{(p(\a) + p(\be))p(\g)}
A^\a_\be B^\g_\de$. In a sense this definition is a contravariant
counterpart of equation (\ref{deftens}). Given matrices $A$, $B$, $C$,
$D$ with operator valued entries, which mutually commute according to
the same rule as $\CL_j$ and $\CL_k$ in eq.\ (\ref{ljlk}), we obtain
for the product of two supertensor products
\beq \label{abcd}
     (A \otimes_s B)(C \otimes_s D) = AC \otimes_s BD \qd.
\eeq
Eq.\ (\ref{gyba}) may be interpreted as defining a graded
Yang-Baxter algebra with $R$-matrix $R$. $\CL_j$ is then its
fundamental representation.

Starting from (\ref{gyba}) we can construct integrable lattice models
as in the non-graded case. Let us briefly recall the construction with
emphasis on the modifications that appear due to the grading. Define
a monodromy matrix $\CT (u,v)$ as an $L$-fold ordered product of
fundamental $L$-matrices,
\beq
     \CT (u,v) = \CL_L (u,v) \dots \CL_1 (u,v) \qd.
\eeq
Due to eq.\ (\ref{homab}) the matrix elements of $\CT(u,v)$ are
homogeneous with parity $p(\CT^\a_\be (u,v)) = p(\a) + p(\be)$. Repeated
application of (\ref{gyba}) and (\ref{abcd}) shows that this monodromy
matrix is a representation of the graded Yang-Baxter algebra,
\beq \label{gtyba}
     \check R(u,v) \left( \CT (u,w) \otimes_s \CT (v,w) \right) =
     \left( \CT (v,w) \otimes_s \CT (u,w) \right) \check R(u,v) \qd.
\eeq
In the non-graded case ($n = 0$) the supertensor product in
(\ref{gtyba}) agrees with the usual tensor product. Multiplying
(\ref{gtyba}) by $\check R^{-1} (u,v)$ and taking the trace of the
whole equation then implies that $[\tr( \CT (u,w)),\tr( \CT (v,w))]
= 0$, and the transfer matrix $\tr( \CT (u,w))$ provides a generating
function of mutually commuting operators, which may take the role of
conserved quantities of an integrable lattice model. For non-trivial
grading the trace has to be replaced by the supertrace, which is
generally defined as $\str( A) = (-1)^{\sum_{j=1}^N p(\a_j)}
A^{\a_1 \dots \a_N}_{\a_1 \dots \a_N}$. Then (\ref{gtyba}) implies that
\beq \label{trans}
     [\str( \CT (u,w)),\str( \CT (v,w))] = 0 \qd,
\eeq
in complete analogy with the non-graded case.

Let us assume that $R(u,v)$ is a regular solution of the Yang-Baxter
equation. This means that there are values $u_0$, $v_0$ of the spectral
parameters such that $R^{\a \be}_{\g \de} (u_0,v_0) = \de^\a_\de
\de^\be_\g$. Then (\ref{defgl}) implies
\beq
     {\CL_j}^\a_\be (u_0,v_0) = (-1)^{p(\a) p(\be)} {e_j}^\a_\be \qd,
\eeq
and we can easily see that the supertrace of the monodromy matrix
at $(u_0,v_0)$ generates a shift by one site,
\bea \label{shift}
     \str(\CT (u_0,v_0)) & = & (-1)^{p(\a)} \CT^\a_\a (u_0,v_0)
          \nn \\
	  & = & (-1)^{\sum_{k=1}^{L-1} p(\be_k)} {e_1}_{\be_L}^{\be_1}
	        {e_2}_{\be_1}^{\be_2} {e_3}_{\be_2}^{\be_3} \dots
		{e_L}_{\be_{L-1}}^{\be_L}
          \nn \\
	  & = & P_{12} P_{23} \dots P_{L-1 L} =: \hat U \qd.
\eea
This implies that $\tau (u) := \ln(\str( \CT (u,v_0)))$ generates a
sequence of local operators \cite{Luescher76} which, as a consequence 
of (\ref{trans}), mutually commute,
\beq \label{tau}
     \tau(u) = \i \hat \Pi + (u - u_0) \hat H + \CO ( (u - u_0)^2 ) \qd.
\eeq
$\hat \Pi$ in this expansion is the momentum operator. On a lattice,
where the minimal possible shift is by one site, and thus $\hat U$
rather than $\hat \Pi$ is the fundamental geometrical operator, some
care is required in the definition of $\hat \Pi$. As was shown in
\cite{GoMu97b} a proper definition may be obtained by setting
$\Pi := - \i \ln ( \hat U) \mod 2 \pi$ and expressing the function
$f(x) = x \mod 2 \pi$ by its Fourier sum. Then $\hat \Pi$ becomes a
polynomial in $\hat U$.
\beq
     \hat \Pi = \Ph \sum_{m=1}^{L-1} \left( \2 +
		 \frac{\hat U^m}{e^{- \i \Ph m} - 1} \right) \qd,
\eeq
where $\Ph = 2 \pi /L$. The first order term $\hat H$ in the expansion
(\ref{tau}) may be interpreted as Hamiltonian. Using (\ref{shift}) it
is obtained as
\beq
     \hat H = \sum_{j=1}^L H_{j j+1}
\eeq
where $H_{L L+1} = H_{L 1}$ and
\beq \label{hdens}
     H_{j j+1} = (-1)^{p(\g)(p(\a) + p(\g))} \,
                 \6_u \left. \check R^{\a \be}_{\g \de} (u,v_0)
		 \right|_{u = u_0} {e_j}_\a^\g {e_{j+1}}_\be^\de \qd.
\eeq
We would like to draw the reader's attention to the following points.
(i) The $R$-matrix $\check R$ in equation (\ref{gyba}) does {\it not}
undergo a modification due to the grading. (ii) The only necessary
compatibility condition which has to be satisfied in order to introduce
graded $L$-matrices is equation (\ref{comp}). As we will see in the
examples below, this is a weak condition. A given $R$-matrix may be
compatible with different gradings, leading to different graded
$L$-matrices.
\subsection*{A first example}
As mentioned above the matrices ${e_j}_\a^\be$ for fixed $j$ form
``local'' representations of gl($m|n$). After summing over all sites,
we obtain a ``global'' representation,
\bea
     E_\a^\be & = & \sum_{j = 1}^L {e_j}_\a^\be \qd, \\[0mm]
     [E_\a^\be,E_\g^\de]_\pm & = &
        \de^\be_\g E_\a^\de -
        (-1)^{(p(\a) + p(\be))(p(\g) + p(\de))} \, \de^\de_\a E_\g^\be
        \qd.
\eea
The $E_\a^\be$ are symmetric in the site indices by construction. Thus
$[P_{j j+1},E_\a^\be] = 0$, and we obtain a gl($m|n$) invariant
Hamiltonian, if we are able to find a solution of the Yang-Baxter
equation (\ref{ybe}), which leads to $H_{j j+1} = P_{j j+1}$ in
equation (\ref{hdens}). Then, comparing (\ref{defp}) and (\ref{hdens}),
$\6_u \left. R^{\a \be}_{\g \de} (u,v_0) \right|_{u = u_0} =
(-1)^{p(\a) p(\be)} \de^\a_\g \de^\be_\de$. Taking into account
regularity, we find the following minimal Ansatz for $R(u,v)$,
\beq \label{rrm}
     R^{\a \be}_{\g \de} (u,v) = \de^\a_\de \de^\be_\g +
          (u - v) (-1)^{p(\a) p(\be)} \de^\a_\g \de^\be_\de \qd,
\eeq
which is indeed a well known rational solution of the Yang-Baxter
equation \cite{Kulish85}.
\subsection*{Global symmetries from local symmetries}
We are going to consider now the general case of symmetries of the
mono\-dromy matrix, which stem from Lie superalgebra invariance of the
$R$-Matrix in a sense to be specified. Choose $x = x^\a_\be e_\a^\be$
homogeneous from gl($m|n$). Let $x_j := x^\a_\be {e_j}_\a^\be$ and
$X := \sum_{j=1}^L x_j$. Define ${\tilde R}^{\a \g}_{\be \de} (u,v)
:= (-1)^{p(\a)p(\g)} R^{\a \g}_{\be \de} (u,v)$. Assume that
${\tilde R} (u,v)$ satisfies the invariance equation
\bea
     \nn
     \lefteqn{
     {\tilde R}^{\a \g}_{\be' \de} (u,v) \, x^{\be'}_\be
     + {\tilde R}^{\a \g}_{\be \de'} (u,v) \, x^{\de'}_\de } \\
     \label{rinv} && =
     (-1)^{p(x)(p(\a') + p(\be))} \, x^\a_{\a'}
     {\tilde R}^{\a' \g}_{\be \de} (u,v) 
     + (-1)^{p(x)(p(\a) + p(\be))} \, x^\g_{\g'}
     {\tilde R}^{\a \g'}_{\be \de} (u,v) . \qd
\eea
From here we can move step by step to the invariance of the transfer
matrix $\str(\CT(u,v))$. Contraction of (\ref{rinv}) with
${e_j}_\g^\de$ yields
\bea
     \nn
     \lefteqn{
     {\CL_j}^\a_{\be'} (u,v) x^{\be '}_\be +
     {\CL_j}^\a_\be (u,v) x_j
     } \\ && =
     (-1)^{p(x)(p(\a') + p(\be))} \, x^\a_{\a'} {\CL_j}^{\a'}_\be (u,v)
     + (-1)^{p(x)(p(\a) + p(\be))} \, x_j {\CL_j}^\a_\be (u,v) \, , \qd
\eea
and it can be shown by induction over $L$ that the monodromy matrix
satisfies
\bea
     \nn \lefteqn{
     \CT^\a_\g (u,v) x^\g_\be + \CT^\a_\be (u,v) X =}\\ && \label{tinv}
        (-1)^{p(x)(p(\g) + p(\be))} \, x^\a_\g \CT^\g_\be (u,v)
        + (-1)^{p(x)(p(\a) + p(\be))} \, X \CT^\a_\be (u,v) \qd.
\eea
We finally take the supertrace of this equation and arrive at
\beq
     [\str(\CT(u,v)),X] = 0 \qd.
\eeq

In many cases the symmetry of the $R$-matrix is evident by construction,
e.g.\ when the $R$-matrix is an intertwiner of representations of
quantum groups. Yet there are examples, as Shastry's $R$-matrix of the
Hubbard model \cite{Shastry88b,OWA87}, where the symmetries are less
obvious \cite{GoMu97b,SUW98}. Moreover, as can be seen by the above
derivation, the symmetries of the transfer matrix are determined by
the symmetries of $\tilde R$ rather than $R$. The symmetries of
$\tilde R$ depend on the choice of the grading.

It may be argued that, in presence of a grading, the matrix $\tilde R$
is more fundamental than $R$, since $\tilde R$ determines the
$L$-matrix, eq.\ (\ref{defgl}), the symmetries of the model and, if it
exists, the semi-classical limit \cite{KuSk82}. Substituting $\tilde R$
into the Yang-Baxter equation (\ref{ybe}), we obtain the so-called
graded Yang-Baxter equation, which equivalently might have been taken
as the starting point of our section on the graded Yang-Baxter algebra.
Since it is the non-graded matrix $\check R$, however, which fixes
the structure of the Yang-Baxter algebra, eq.\ (\ref{gyba}), we stood
away from this point of view.

\subsection*{Representations in terms of fermions}
In this section we shall explain how the various graded objects, which
have been introduced so far, can be expressed in terms of
Fermi operators. To begin with, consider spinless fermions on a ring of
$L$ lattice sites,
\beq \label{antic}
     \{c_j,c_k\} = \{c_j^\dagger,c_k^\dagger\} = 0 \qd, \qd
     \{c_j,c_k^\dagger\} = \de_{jk} \qd, \qd j, k = 1, \dots , L \qd.
\eeq
Locally there are two states, every site is either occupied by a fermion
or it is empty. Slightly deviating from the usual language we may say
that $c_j$ and $c_j^\dagger$ annihilate or create the occupied state.
Let us define a pair ${a_j}_1$, ${a_j}_1^\dagger$ of (trivial)
annihilation and creation operators of the unoccupied state by setting
${a_j}_1 = {a_j}_1^\dagger = 1$, and let us write ${a_j}_2 = c_j$,
${a_j}_2^\dagger = c_j^\dagger$. Let $n_j = c_j^\dagger c_j$ denote
the density operator.
The operators
\beq
     {Y_j}_\a^\be = {a_j}_\a^\dagger (1 - n_j) {a_j}_\be
\eeq
are then obviously local projection operators, i.e.\ they satisfy
\beq \label{xpro}
     {Y_j}_\a^\be {Y_j}_\g^\de = \de_\g^\be {Y_j}_\a^\de \qd.
\eeq
They carry parity, induced by the anticommutation rule for the
Fermi operators. Let $j \ne k$. Then ${Y_j}_\a^\be$ and ${Y_k}_\g^\de$
anticommute, if both are build up of an odd number of Fermi operators,
and commute in all other cases. This fact can be expressed as follows.
Let $p(1) = 0$, $p(2) = 1$ and $p({Y_j}_\a^\be) = p(\a) + p(\be)$. Then
${Y_j}_\a^\be$ is odd (contains an odd number of Fermi operators), if
$p({Y_j}_\a^\be) = 1$, and even, if $p({Y_j}_\a^\be) = 0$. The
commutation rules for the projectors ${Y_j}_\a^\be$ are thus
\beq \label{xcom}
     {Y_j}_\a^\be {Y_k}_\g^\de = (-1)^{(p(\a) + p(\be))(p(\g) + p(\de))}
        {Y_k}_\g^\de {Y_j}_\a^\be \qd,
\eeq
As a memorizing scheme for the projection operators let us combine
them into the matrix
$(Y_j)^\a_\be = {Y_j}_\a^\be$,
\beq \label{yj1}
     Y_j =
        \left(
	\begin{array}{cc} 1 - n_j & c_j \\ c_j^\dagger & n_j \end{array}
        \right) \qd.
\eeq
(\ref{xpro}) and (\ref{xcom}) are representations of equations
(\ref{coantico}) and (\ref{samesite}) in the particular case $m = n =1$.
Since all our considerations in the previous sections entirely relied
on eqs.\ (\ref{coantico}) and (\ref{samesite}), we may simply replace
${e_j}_\a^\be$ by ${Y_j}_\a^\be$ in eqs.\ (\ref{defp}), (\ref{defgl})
and (\ref{hdens}) to obtain fermionic representations of the permutation
operator, the $L$-matrix and the Hamiltonian. The permutation operator
(for $j \ne k$) becomes
\bea \nn
     P_{jk} & = & {Y_j}_1^1 {Y_k}_1^1 + {Y_j}_2^1 {Y_k}_1^2 -
        {Y_j}_1^2 {Y_k}_2^1 - {Y_j}_2^2 {Y_k}_2^2 \\ \label{pjk1}
	    & = & 1 - (c_j^\dagger - c_k^\dagger)(c_j - c_k) \qd.
\eea

Fermionic representations compatible with arbitrary grading can be
constructed by considering several species of fermions and graded
products of projection operators. We shall explain this for the case
of two species first. This is the most interesting case for
applications, since we may interpret the two species as up- and
down-spin electrons. We have to attach a spin index to the
Fermi operators, $c_j \rightarrow c_{j \s}$, $\s = \auf, \ab$,
$\{c_{j \s},c_{k \tau}^\dagger\} = \de_{jk} \de_{\s \tau}$.
Accordingly, there are two species of projection operators,
${Y_j}_\a^\be \rightarrow {Y_j^\s}_\a^\be$,
\beq
     {Y_j^\auf}_\a^\be {Y_j^\ab}_\g^\de = 
        (-1)^{(p(\a) + p(\be))(p(\g) + p(\de))}
        {Y_j^\ab}_\g^\de {Y_j^\auf}_\a^\be  \qd.
\eeq
Let us define projection operators for electrons by the tensor products
\beq
     {Y_j}_{\a \g}^{\be \de} = (-1)^{(p(\a) + p(\be))p(\g)}
        {Y_j^\ab}_\a^\be {Y_j^\auf}_\g^\de =
	\left( Y_j^\ab \otimes_s Y_j^\auf \right)^{\a \g}_{\be \de} \qd.
\eeq
Then
\beq
     {Y_j}_{\a \g}^{\be \de} {Y_j}_{\a' \g'}^{\be' \de'} =
        \de_{\a'}^\be \de_{\g'}^\de {Y_j}_{\a \g}^{\be' \de'} \qd.
\eeq
${Y_j}_{\a \g}^{\be \de}$ inherits the parity from ${Y_j^\ab}_\a^\be$
and ${Y_j^\auf}_\g^\de$. The number of Fermi operators contained in
${Y_j}_{\a \g}^{\be \de}$ is the sum of the number of Fermi operators
in ${Y_j^\ab}_\a^\be$ and ${Y_j^\auf}_\g^\de$. Hence
$p({Y_j}_{\a \g}^{\be \de}) = p({Y_j^\ab}_\a^\be) + p({Y_j^\auf}_\g^\de)
= p(\a) + \dots + p(\de)$, and the analog of (\ref{xcom}) holds for
${Y_j}_{\a \g}^{\be \de}$, too. Again we present all projection
operators in form of a matrix $(Y_j)^{\a \g}_{\be \de} =
{Y_j}_{\a \g}^{\be \de}$,
\bea
     Y_j & = & Y_j^\ab \otimes_s Y_j^\auf \nn \\ \label{yj}
        & = & \left(
        \begin{array}{cccc}
	   (1 - n_{j \ab})(1 - n_{j \auf}) &
	   (1 - n_{j \ab}) c_{j \auf} &
	   c_{j \ab} (1 - n_{j \auf}) &
	   c_{j \ab} c_{j \auf} \\
	   (1 - n_{j \ab}) c_{j \auf}^\dagger &
	   (1 - n_{j \ab}) n_{j \auf} &
	   - c_{j \ab} c_{j \auf}^\dagger &
	   - c_{j \ab} n_{j \auf} \\
	   c_{j \ab}^\dagger (1 - n_{j \auf}) &
	   c_{j \ab}^\dagger c_{j \auf} &
	   n_{j \ab} (1 - n_{j \auf}) &
	   n_{j \ab} c_{j \auf} \\
	   - c_{j \ab}^\dagger c_{j \auf}^\dagger &
	   - c_{j \ab}^\dagger n_{j \auf} &
	   n_{j \ab} c_{j \auf}^\dagger &
	   n_{j \ab} n_{j \auf}
        \end{array} \right) \,. \qd
\eea
Here we used the standard ordering of matrix elements of tensor
products, corresponding to a renumbering $(11) \rightarrow 1$,
$(12) \rightarrow 2$, $(21) \rightarrow 3$, $(22) \rightarrow 4$.
Within this convention ${Y_j}_{\a \g}^{\be \de}$ is replaced by
${Y_j}_\a^\be$, $\a, \be = 1, \dots, 4$, which then satisfies
(\ref{xpro}) and (\ref{xcom}) with grading $p(1) = p(4) = 0$,
$p(2) = p(3) = 1$.

The permutation operator of electrons turns out to be a product of
permutation operators of up-and down-spin electrons,
\bea \label{pgl22}
     P_{jk} & = & (-1)^{p(\be) + p(\de)} {Y_j}_{\a \g}^{\be \de}
                  {Y_k}_{\be \de}^{\a \g} \nn \\
            & = & (-1)^{p(\be) + p (\de) + (p(\a) + p(\be))
	                                      (p(\g) + p(\de))}
                  {Y_j^\ab}_\a^\be {Y_j^\auf}_\g^\de 
                  {Y_k^\ab}^\a_\be {Y_k^\auf}^\g_\de \nn \\
            & = & (-1)^{p(\be)} {Y_j^\ab}_\a^\be {Y_k^\ab}^\a_\be
	          (-1)^{p(\de)} {Y_j^\auf}_\g^\de {Y_k^\auf}^\g_\de
              = P_{jk}^\ab P_{jk}^\auf
              = P_{jk}^\auf P_{jk}^\ab \qd.
\eea

So far we have considered the case of spinless fermions with
two-di\-men\-sional local space of states and grading $m = n =1$, and
the case of electrons with four-dimensional space of states and grading
$m = n = 2$. There are four different possibilities to realize
(\ref{coantico}) and (\ref{samesite}) in case of a three-dimensional
local space of states, $m + n = 3$. They can be obtained by deleting
the $\a$'s row and column of the matrix $Y_j$ in eq.\ (\ref{yj}),
$\a = 1, 2, 3, 4$. (\ref{xpro}) and (\ref{xcom}) remain valid, since
the operators ${Y_j}_\a^\be$ are projectors.

As an example let us consider the case where the fourth row and column
of $Y_j$, eq. (\ref{yj}), are deleted. The local Hilbert space is then
spanned by the three states $|0\>$, $c_{j \auf}^\dagger |0\>$,
$c_{j \ab}^\dagger |0\>$, i.e.\ double occupancy is now excluded. The
operator
\beq
     \sum_{\a=1}^3 {Y_j}_\a^\a = 1 - {Y_j}_4^4 =
        1 - n_{j \auf} n_{j \ab}
\eeq
projects the local Hilbert space of electrons onto the space with no
double occupancy. The global projection operator for a chain of $L$
sites is given by the product
\beq \label{notwo}
     \D = \prod_{j=1}^L (1 - n_{j \auf} n_{j \ab}) \qd.
\eeq
The permutation operator $P_{jk}$ is again given by eq.\ (\ref{defp})
with ${Y_j}_\a^\be$ replacing ${e_j}_\a^\be$. Summation is now over
three values, $\a, \be = 1, 2, 3$, and the grading is $p(1) = 0$,
$p(2) = p(3) = 1$. An elegant way of taking into account the
simplifications arising from the restriction to the Hilbert space with
with no double occupancy is to consider $P_{jk} \D$ instead of
$P_{jk}$. Since $n_{j \auf} n_{j \ab} \D = 0$, we obtain
\beq \label{ptj}
     P_{jk} \D = \D (c_{j \s}^\dagger c_{k \s} +
        c_{k \s}^\dagger c_{j \s}) \D -
	2 (S_j^a S_k^a - \tst{\4} n_j n_k) \D + (1 - n_j - n_k) \D \qd.
\eeq
Here we have introduced the electron density $n_j = n_{j \auf} +
n_{j \ab}$ and the spin densities
\beq \label{defs}
     S_j^a = \tst{\2} \s_{\a \be}^a c_{j \a}^\dagger c_{j \be} \qd.
\eeq
The $\s^a$, $a = x, y, z$, are the Pauli matrices, and we identify
1 with $\auf$ an 2 with $\ab$ in the summation over $\a$ and $\be$.
The spin densities can alternatively be written as
\beq
     S_j^a = \tst{\2} \left( \s^a \right)^\a_\be
                {Y_j}_{\a + 1}^{\be + 1} \qd.
\eeq
Using
\beq
        \left( \s^a \otimes \s^a \right)^{\a \g}_{\be \de} =
        2 \de^\a_\de \de^\g_\be - \de^\a_\be \de^\g_\de \qd,
\eeq
we obtain
\beq
     \left( {Y_j}_1^1 {Y_k}_1^1-
        \sum_{\a, \be = 2}^3 {Y_j}_\a^\be {Y_k}_\be^\a \right) \D =
	\left( -2 (S_j^a S_k^a - \tst{\4} n_j n_k) + 1 - n_j - n_k
	\right) \D \qd,
\eeq
which gives the second and third term on the right hand side of
(\ref{ptj}). Note that the permutation operator $P_{jk}$ in eq.\
(\ref{ptj}) is no longer a product like in eq.\ (\ref{pgl22}).

It should be clear by now, how to generalize the above considerations
to an arbitrary number of species of fermions. In the case of $N$
species we may define
\beq \label{multipro}
     {Y_j}_{\a_1 \dots \a_N}^{\be_1 \dots \be_N} =
        \left( Y_j^N \otimes_s \dots \otimes_s Y_j^1
	\right)^{\a_1 \dots \a_N}_{\be_1 \dots \be_N} \qd.
\eeq
Then
\bea
     {Y_j}_{\a_1 \dots \a_N}^{\be_1 \dots \be_N}
     {Y_j}_{\g_1 \dots \g_N}^{\de_1 \dots \de_N} & = &
     \de_{\g_1}^{\be_1} \dots \de_{\g_N}^{\be_N}
     {Y_j}_{\a_1 \dots \a_N}^{\de_1 \dots \de_N} \qd, \\
     {Y_j}_{\a_1 \dots \a_N}^{\be_1 \dots \be_N}
     {Y_k}_{\g_1 \dots \g_N}^{\de_1 \dots \de_N} & = &
     (-1)^{\sum_{j,k = 1}^N (p(\a_j) + p(\be_j))(p(\g_k) + p(\de_k))}
     {Y_k}_{\g_1 \dots \g_N}^{\de_1 \dots \de_N}
     {Y_j}_{\a_1 \dots \a_N}^{\be_1 \dots \be_N} \, , \qd
\eea
which can be shown by induction over the number of species. Here the
grading is $m = n = 2^{N -1}$. The most general case is obtained by
deleting rows and columns from $Y_j$, eq.\ (\ref{multipro}), in
analogy to the example considered above.

Let us note that the operators ${Y_j}_\a^\be$ for two species of
fermions appear under the name Hubbard projection operators in the
literature.
\subsection*{Gauge transformations}
The canonical anticommutation relations (\ref{antic}) are invariant
under local gauge transformations
\beq
     c_j \qd \rightarrow \qd \hat c_j = e^{\i \ph_j} c_j \qd, \qd
        \ph_j \in [0,2\p] \qd.
\eeq
Thus all our formulae remain correct, if we replace $c_j$ by $\hat c_j$.
The local gauge transformation induces a transformation of the matrix
$Y_j$ of projection operators,
\beq \label{xgauge}
     Y_j \qd \rightarrow \qd \hat Y_j =
        e^{\i \ph_j \s^z/2} Y_j e^{- \i \ph_j \s^z/2} =
	e^{- \i \ph_j n_j} Y_j e^{\i \ph_j n_j} \qd.
\eeq
Here $\s^z$ is a Pauli matrix. In the case of $N$ species of fermions
there are $N$ gauge parameters $\ph_j^\s$, $\s = 1, \dots, N$. Since
$e^{\i \ph_j^\s \s^z/2}$ is diagonal, the form (\ref{xgauge}) of the
transformation rule for $Y_j$ carries over to the case of $N$ species.
Let
\beq
     G_j = e^{\i \ph_j^N \s^z/2} \otimes \dots \otimes
        e^{\i \ph_j^1 \s^z/2} \qd.
\eeq
Then
\beq
     Y_j \qd \rightarrow \qd \hat Y_j = G_j Y_j G_j^\dagger \qd.
\eeq
For the sake of simplicity we assume in the following that we
are dealing with two species of fermions. Note, however, that the
following considerations also apply in the most general case, where
we would have to deal with a sub-matrix of $Y_j$, eq.\ (\ref{multipro}).
The $L$-matrix (\ref{defgl}) behaves under gauge transformations as
\beq
     \hat{\CL_j}^\a_\be = {G_j^\dagger}^\a_{\a'} (-1)^{p(\a')p(\g')}
        \hat{R_j}^{\a' \g'}_{\be' \de'} (u,v) {Y_j}_{\g'}^{\de'}
	{G_j}^{\be'}_\be \qd.
\eeq
Here we defined
\beq \label{gauger}
     \hat R_j (u,v) = (G_j \otimes G_j) R(u,v)
                      (G_j^\dagger \otimes G_j^\dagger) \qd.
\eeq

Let us consider two implications of gauge invariance of the $R$-matrix.

(i) Global gauge transformations: Assume that $G_j = G$, $j = 1, \dots
L$, and that $\hat R (u,v) = R (u,v)$, say, for arbitrary $\ph^\auf$,
$\ph^\ab$ ($\auf = 1$, $\ab = 2$). Taking the derivative of
(\ref{gauger}) with respect to $\ph^\auf$ at $\ph^\auf = \ph^\ab = 0$
yields
\beq
     [R(u,v),I_2 \otimes \s^z \otimes I_4 +
        I_4 \otimes I_2 \otimes \s^z] = 0 \qd.
\eeq
Since $I_2 \otimes \s^z \otimes I_4$ and $I_4 \otimes I_2 \otimes \s^z$
are diagonal, we may replace $R (u,v)$ in this equation by
$\tilde R (u,v)$. Then (\ref{rinv}) is satisfied with $x = I_2 \otimes
\s^z$. Thus the transfer matrix commutes with
\beq
     X = \sum_{j=1}^L \left( {Y_j}^1_1 - {Y_j}^2_2 + {Y_j}^3_3
                            - {Y_j}^4_4 \right) =
         \sum_{j=1}^L (1 - 2 n_{j \auf}) \qd,
\eeq
which means that the number of up-spin electrons is conserved. A
similar statement is easily verified for the number of down-spin
electrons or, in the weaker case, when (\ref{gauger}) is only satisfied
for $\ph^\auf = \ph^\ab$, for the total number of electrons.

(ii) Local gauge transformations: Assume that $G_j = (G)^j$, where
$G$ is a fixed matrix as in the example above, and that $R(u,v)$ is
invariant under $G$. Then
\beq
     \hat \CL_j (u,v) = \left( G^\dagger \right)^j \CL_j (u,v) (G)^j
                        \qd,
\eeq
and the transformed monodromy matrix becomes a simple expression in
terms of the original $L$-matrices,
\beq
     \hat \CT (u,v) = \left( G^\dagger \right)^L \CL_L (u,v) G
        \CL_{L - 1} (u,v) G \dots \CL_1 (u,v) G \qd.
\eeq
Transformations of this type can be used to introduce a phase factor
into a typical nearest-neighbor hopping term as it appears in the
Hubbard Hamiltonian. If the phase factor is just $-1$, then the
hopping term changes sign. The factor $(G^\dagger)^L$ generally modifies
the transfer matrix $\str(\hat \CT (u,v))$. It leads to a twist of the
periodic boundary conditions. Note however, that it may happen for
certain values of $L$, e.g.\ if $L$ is divisible by 2 or by 4, that
$(G^\dagger)^L = 1$.

Gauge transformations of the type considered modify the shift operator.
Using the right hand side of (\ref{xgauge}) in eq.\ (\ref{shift}) we
can easily see that
\beq \label{shiftgauge}
     \hat U \qd \rightarrow \qd e^{\i L \ph^\s n_{1 \s} - \i \ph^\s
        \sum_{j=1}^L n_{j \s}} \, \hat U \qd,
\eeq
where summation over $\s = \auf, \ab$ is implied. The first term in
the exponent of (\ref{shiftgauge}) is related to a twist of boundary
conditions.

At this point will will not go into further detail, because we think
that more detailed considerations are only sensible in the context of
concrete models. Let us only remark that global symmetries of a given
model may be affected by the exponent in eq.\ (\ref{shiftgauge}), even
if there is no twist, i.e.\ even if the first term is equal to zero
modulo $2\p$.

In the context of integrable fermionic models the term gauge
transformation has, unfortunately, two different meanings, which should
not be confused. Besides the meaning discussed above it is also
used for transformations on the $R$-matrix of the form $R \rightarrow
(G \otimes G) R (G^{-1} \otimes G^{-1})$, where $G$ is not necessarily
a diagonal matrix. Transformations of this form leave the Yang-Baxter
equation (\ref{ybe}) invariant.
\subsection*{More examples}
To illustrate the formal discussions of the preceding sections let us
further elaborate on the models with rational $R$-matrices (\ref{rrm}).
These models have been studied exhaustively in the literature. They
appeared first as lattice gas models of Lai \cite{Lai74} and Sutherland
\cite{Sutherland75} and were solved by coordinate Bethe ansatz. Later
Kulish studied them within the framework of the graded quantum inverse
scattering method and obtained spectrum and eigenstates by means of
the nested algebraic Bethe ansatz \cite{Kulish85}. Still Kulish did
not write down any Hamiltonian density in terms of Fermi operators.
This was first accomplished by Schlottmann for the gl(1$|$2) invariant
case \cite{Schlottmann87}. The Hamiltonian in fermionic representation
is the Hamiltonian of the super symmetric $t$-$J$-model. Schlottmann
solved it again by means of the coordinate Bethe ansatz. However, he
was not aware of the underlying algebraic structure. The underlying
algebraic structure was successively unravelled by different authors
\cite{BaBl90,Sarkar91,EsKo92} leading to a solution of the super
symmetric $t$-$J$-model by nested algebraic Bethe ansatz
\cite{EsKo92,FoKa93}. The gl(2$|$2) invariant model with the Hamiltonian
in fermionic representation was studied by E{\ss}ler et al.\ %
\cite{EKS92,EKS94}.

Let us write down Hamiltonian, $R$-matrix, $L$-matrix and generators
of symmetries for the gl(1$|$1) invariant case using the fermionic
representation (\ref{yj1}). The $R$-matrix follows from (\ref{rrm})
with $p(1) = 0$, $p(2) = 1$,
\beq
     \check R (u,v) = \left( \begin{array}{cccc}
                      1 + u - v &&&\\ &1& u - v &\\ & u - v &1&\\
		      &&& 1 - u + v \end{array} \right) \qd.
\eeq
The $L$-matrix is obtained from (\ref{defgl}),
\bea
     \CL_j (u,v) & = & \left( \begin{array}{cc}
                       u - v + {e_j}_1^1 & {e_j}_2^1\\
		       {e_j}_1^2 & u - v - {e_j}_2^2 \end{array} \right)
		       \\ & \rightarrow &
                       \left( \begin{array}{cc}
                       u - v + 1 - n_j & c_j^\dagger \\
		       c_j & u - v - n_j \end{array} \right) \qd.
\eea
As Hamiltonian density we may choose
\bea \label{hgl11}
     H_{j j+1} & = & - P_{j j+1} = (c_j^\dagger - c_{j+1}^\dagger)
                                   (c_j - c_{j+1}) - 1 \nn \\
               & = & - (c_j^\dagger c_{j+1} + c_{j+1}^\dagger c_j)
	             + n_j + n_{j+1} - 1
\eea
(cf (\ref{pjk1})), which is the Hamiltonian density of a system of
free fermions. The odd generators of symmetries are
\bea
     E_2^1 & = & \sum_{j=1}^L {e_j}_2^1 \qd \rightarrow \qd
                 \sum_{j=1}^L c_j^\dagger \qd, \\
     E_1^2 & = & \sum_{j=1}^L {e_j}_1^2 \qd \rightarrow \qd
                 \sum_{j=1}^L c_j \qd.
\eea
From the even generators we can construct only one non-trivial
combination,
\beq
     \tst{\2} (E_2^2 - E_1^1 + L) \qd \rightarrow \qd
         \hat N = \sum_{j=1}^L n_j \qd.
\eeq
Of course, being a model of free fermions, (\ref{hgl11}) is trivial in
the sense that the corresponding Hamiltonian can be diagonalized by
Fourier transform. We wrote down the above formulae in order to
illustrate the general formalism. The reader can easily repeat all the
steps for the case of the super symmetric $t$-$J$-model, where the
Hamiltonian density is given by $H_{j j+1} = - P_{j j+1}$ with
$P_{j j+1}$ according to eq.\ (\ref{ptj}).

Let us indicate the Hamiltonian density of the gl(2$|$2) invariant
model \cite{EKS92,EKS94} using the fermionic representation (\ref{yj}).
(\ref{rrm}) and (\ref{hdens}) imply that
\bea \label{hgl22}
     H_{j j+1} & = & - P_{j j+1} = - P_{j j+1}^\auf P_{j j+1}^\ab \\
               & = & - (1 - (c_{j \auf}^\dagger - c_{j+1 \auf}^\dagger)
	                (c_{j \auf} - c_{j+1 \auf}))
                       (1 - (c_{j \ab}^\dagger - c_{j+1 \ab}^\dagger)
	                (c_{j \ab} - c_{j+1 \ab})) . \qd \nn
\eea
Here the minus sign on the right hand side has been chosen in order
to meet the usual conventions. If we open the brackets in (\ref{hgl22}),
we obtain the rather voluminous expression $H_{j j+1}^0$ of E{\ss}ler,
Korepin and Schoutens (cf e.g.\ eq.\ (8) in \cite{EKS92}), which we
do not repeat here due to space limitations. Since
$[P_{j j+1},E_\a^\be] = 0$, $\a, \be = 1, 2, 3, 4$, we may add,
for instance, $U E_4^4$, $U$ real, to the Hamiltonian obtainable from
(\ref{hgl22}) without spoiling integrability (see \cite{EKS92,EKS94}).
This amounts to adding a Hubbard interaction $U n_{j \auf} n_{j \ab}$
to the Hamiltonian density. Note that the factorized form (\ref{hgl22})
of $H_{j j+1}$ appears to be new. 

Here is a more recent example. Maassarani and Mathieu constructed an
``su($N$) analogue'' of the XX-model and found the corresponding
$R$-matrix \cite{MaMa97a}. We will show that a certain fermionic
representation of the ``su(3) version'' of this $R$-matrix generates
the Hamiltonian of the Hubbard model at infinite coupling. Let us
write the $R$-matrix of Maassarani and Mathieu in the form
\bea \label{rmama}
     \nn
     \lefteqn{\check R (u,v|\de) = \sin(u - v) \sum_{\a = 2}^N
        \left( e^{\i \de} e^1_\a \otimes e^\a_1 + e^{- \i \de}
	       e^\a_1 \otimes e^1_\a \right)}\\ &&
        + \sum_{\a = 2}^N \left( e^1_1 \otimes e^\a_\a +
	                         e^\a_\a \otimes e^1_1 \right)
        + \cos(u - v) \left( e^1_1 \otimes e^1_1 + \sum_{\a, \be = 2}^N
	                     e^\a_\a \otimes e^\be_\be \right) \, . \qd
\eea
Note that $\check R$ depends on an additional free parameter $\de$. In
order to meet our conventions we modified the original $R$-matrix of
Maassarani and Mathieu by a gauge transformation $G$, $\check R
\rightarrow (G \otimes G) \check R (G^{-1} \otimes G^{-1})$,
$G e_\a^\be G^{-1} = e_{(\a + 1) \smod N}^{(\be + 1) \smod N}$. Clearly,
$\check R$ is regular. Let us now restrict ourselves to the case
$N = 3$. Then $\check R$ is compatible with the grading $p(1) = 0$,
$p(2) = p(3) = 1$. Let us consider the fermionic representation
which is obtained from $Y_j$, eq.\ (\ref{yj}), by deleting the fourth
row and column. Using (\ref{hdens}) we obtain the Hamiltonian density
\beq
     H_{j j+1} = \sum_{\s = \auf, \ab}
                 (e^{\i \de} c_{j \s}^\dagger c_{j+1, \s} +
                 e^{- \i \de} c_{j+1, \s}^\dagger c_{j \s})
		 (1 - n_{j, -\s})(1 - n_{j+1, -\s}) \qd.
\eeq
We see that $\de$ is connected to local gauge transformations. Setting
$\de = \p$, $H_{j j+1}$ turns into the Hamiltonian density of the
Hubbard model at infinite coupling, which is the same as the
restricted hopping part of the super symmetric $t-J$-Hamiltonian.
Hence the Hamiltonian can be written in the more familiar form
\beq \label{hhinf}
     H = - \sum_{j=1}^L \D (c_{j \s}^\dagger c_{j+1, \s} +
                            c_{j+1, \s}^\dagger c_{j \s}) \D \qd,
\eeq
where $\D$, eq.\ (\ref{notwo}), is the projection operator which
excludes double occupancy.  Of course, (\ref{hhinf}) only makes sense,
if the number of particles is less than $L$.
                  
\subsection*{Algebraic Bethe ansatz for the Hubbard model in the
infinite coupling limit}

We can now apply the algebraic Bethe ansatz \cite{KBIBo} to construct
eigenvectors of the Hamiltonian (\ref{hhinf}). We shall regard equation
(\ref{gtyba}) as a set of algebraic relations between the elements of
the monodromy matrix $\CT (u,w)$.  Since $\CT (u,w)$ is a function of
$u-w$ in this case, we shall simply write $\CT(u,w)=\CT(u-w)$ hereafter.
Following Maassarani and Mathieu \cite{MaMa97a}, we shall use the
notation 
\beq
     \CT(u)  =  \left( \begin{array}{ccc}
                   S(u) & C_{1}(u) & C_{2}(u) \\
                   B_{1}(u) & t_1^1(u) & t_2^1(u) \\
		   B_{2}(u) & t_1^2(u) & t_2^2(u)
		\end{array} \right) \qd.
\eeq
Let $|0\>$ denote the vacuum state defined by $c_j|0\> = 0$. Using the
explicit form of the $L$-matrix, which follows from (\ref{defgl}),
(\ref{rmama}),
\beq
 \CL_{j}(u)  =  \left(
        \begin{array}{ccc}
           \sst{\cos u\, {Y_j}_1^1 - \sin u\, 
           ({Y_j}_2^2 + {Y_j}_3^3)} & 
           \sst{{Y_j}^1_2} &
	   \sst{{Y_j}^1_3} \\
           \sst{{Y_j}^2_1} &
           \sst{- \sin u\, {Y_j}_1^1 - \cos u\, {Y_j}_2^2} &
           \sst{- \cos u\, {Y_j}_3^2} \\
           \sst{{Y_j}^3_1} &
           \sst{- \cos u\, {Y_j}_2^3} &
           \sst{- \sin u\, {Y_j}_1^1 - \cos u\, {Y_j}_3^3}
        \end{array} \right) \, ,
\eeq
we can easily see that 
\beq
 \CT(u)|0\rangle  =  \left(
        \begin{array}{ccc}
        s(u)|0\> & * & * \\
        0 & t(u)|0\> & 0 \\
        0 & 0 & t(u)|0\>
        \end{array} \right) \qd ,
\eeq
where $s(u) = \cos^L (u)$, $t(u) = (-1)^L \sin^L (u)$. This equation
indicates that the vacuum $|0\>$ is an eigenstate of the transfer
matrix. Following the procedure of the nested algebraic Bethe ansatz
\cite{Kulish85,EsKo92}, we assume the state
\beq \label{abastate}
     |\la_1,\cdots,\la_p\> = F^{a_1 \dots a_p} C_{a_1}(\la_1)
                             \dots C_{a_p}(\la_p)|0\>
\eeq
to be an eigenstate of the transfer matrix. Here summation over
$a_j = 1, 2$ is understood.

The commutation rules between the elements of the monodromy matrix can 
be extracted from (\ref{gtyba}),
\bea \label{cccom}
     C_a(u)C_b(v) & = & C_a(v)C_b(u) \qd, \\
     S(u)C_a(v) & = & \cot(u - v) C_a(v)S(u) - \frac{C_a(u)S(v)}
                      {\sin(u - v)} \qd, \\
     t_a^a(u)C_b(v) & = & \cot(u - v) C_a(v)t_b^a(u) -
                          \frac{C_a(u)t_b^a(v)} {\sin(u - v)} \qd.
\eea
Comparing these commutation rules with the ungraded case in
\cite{MaMa97a}, we notice that there appear extra minus signs when
commuting $t_a^a(u)$ and $C_b(v)$. We use the above relations to
calculate the action of the transfer matrix $\str(\CT(u))$ on the state 
$|\la_1,\dots, \la_p\>$. The resulting terms are classified as wanted
terms and unwanted terms, respectively. The wanted terms are
\bea
     \lefteqn{
     \prod_{j=1}^p \cot(u - \la_j) \{ s(u) |\la_1,\dots,\la_p\>}
        \nn \\ && \qqd
        - t(u) (\tau^{(p)} F)^{a_1 \dots a_p} C_{a_1} (\la_1) \dots
	C_{a_p} (\la_p) |0\> \} \qd.
\eea
Here $\tau^{(p)}$ is the shift operator on a $2^L$ dimensional
auxiliary space, which may be realized as the space of states of a
$p$-site spin one-half chain. Its matrix elements are given as 
\beq
     {\tau^{(p)}}_{a_1 \dots a_p}^{b_1 \dots b_p} =
        \de_{a_p}^{b_1} \de_{a_1}^{b_2} \dots \de_{a_{p-1}}^{b_p} \qd.
\eeq
The wanted terms are proportional to $|\la_1,\dots,\la_p\>$, if $F$
is an eigenvector of the shift operator $\tau^{(p)}$ with eigenvalue
$\La^{(p)}$. On the other hand, in oder for the unwanted terms to
cancel each other, we have the condition
\beq
     \tau^{(p)} F = (-1)^L \cot^L (\la_j) F \qd.
\eeq
The spectrum of $\tau^{(p)}$ is fixed by the condition
$(\tau^{(p)})^p = 1$. This condition implies that the spectrum consists
of powers of the $p$-th root of unity (cf appendix B of
\cite{GoMu97b}). The eigenvalue $\La(u)$ of the transfer matrix
$\str(\CT(u))$ is thus given as
\beq
     \La(u) = \left( s(u) - t(u) e^{\frac{\i 2\p l}{p}} \right)
              \prod_{j=1}^p \cot(u - \la_j) \qd,
\eeq
where $\la_j$, $j = 1, \dots, p$, is a solution of the Bethe ansatz
equations
\beq \label{baeqn}
     (-1)^L \cot^L (\la_j) = e^{\frac{\i 2\p l}{p}} \qd, \qd
                             l = 0, \dots, p - 1 \qd.
\eeq

The only remaining problem is the explicit construction of eigenvectors
of the shift operator $\tau^{(p)}$. This is, of course, a highly
degenerate problem. We may use any integrable spin one-half chain
with regular $R$-matrix to solve this problem. A particularly simple
choice is the XX spin chain. Its $R$-matrix $\check R$ is given by
eq.\ (\ref{rmama}) with $N = 2$ and $\de = 0$. The $L$-matrix in
(non-graded) spin representation is
\beq
     L_j(u) = \left( \begin{array}{cc}
                 \cos(u) {e_j}_1^1 + \sin(u) {e_j}_2^2 & {e_j}_2^1 \\
		 {e_j}_1^2 & \sin(u) {e_j}_1^1 + \cos(u) {e_j}_2^2
              \end{array} \right) \qd.
\eeq
Let us use the notation
\beq
     T(u) = \left( \begin{array}{cc}
               A(u) & B(u) \\ C(u) & D(u) \end{array} \right)
\eeq
for the monodromy matrix $T(u) = L_p(u) \dots L_1(u)$. With the choice
$|0) = {1 \choose 0}^{\otimes p}$ of auxiliary vacuum, we obtain
\beq
     T(u)|0) = \left( \begin{array}{cc}
                  \cos^p(u)|0) & * \\ 0 & \sin^p(u)|0)
	       \end{array} \right) \qd,
\eeq
which shows that the auxiliary vacuum is an eigenstate of the
transfer matrix $\tr(T(u))$. To construct the eigenstates of
$\tr(T(u))$ we need the commutation relations
\bea \label{bb}
     [B(u),B(v)] & = & 0 \qd, \\
     A(u)B(v) & = & - \cot(u - v) B(v)A(u) + \frac{B(u)A(v)}
                      {\sin(u - v)} \qd, \\
     D(u)B(v) & = & \cot(u - v) B(v)D(u) -
                          \frac{B(u)D(v)} {\sin(u - v)} \qd,
\eea
which are part of the Yang-Baxter algebra (\ref{gtyba}). Proceeding as
above, we consider the action of the transfer matrix $\tr(T(u)) = A(u)
+ D(u)$ on the state
\beq
     |\m_1, \dots, \m_s) = B(\m_1) \dots B(\m_s)|0) \qd.
\eeq
It turns out that $|\m_1, \dots, \m_s)$ is an eigenstate of $\tr(T(u))$
with eigenvalue
\beq \label{tfxx}
     \La^{(p)} (u) = \left\{ (-1)^s \cos^p(u) + \sin^p(u) \right\}
                     \prod_{j=1}^s \cot(u - \m_j) \qd,
\eeq
if the parameters $\m_j$, $j = 1, \dots, s$, satisfy the equation
\beq \label{xxba}
     \cot^p (\m_j) = (-1)^{s+1} \qd.
\eeq
Eq.\ (\ref{tfxx}) implies in particular that
\beq \label{tfxxzero}
     \La^{(p)} (0) = \prod_{j=1}^s \cot (\m_j) \qd.
\eeq
In order to introduce a convenient basis let us use the operators
${e_j}_a^1$. ${e_j}_a^1$ applied to the auxiliary vacuum $|0)$ places
an $\auf$-spin at site $j$, if $a = 1$, and a $\ab$-spin, if $a = 2$.
The states ${e_1}_{a_1}^1 \dots {e_p}_{a_p}^1|0)$ form an orthonormal
basis of our auxiliary Hilbert space, since ${e_j}_1^b {e_j}_a^1 =
\de_a^b {e_j}_1^1$ and ${e_j}_1^1|0) = |0)$. The components of the
shift operator $\hat U = \tr(T(0))$ with respects to this basis are
\bea \nn
     \lefteqn{(0|{e_p}_1^{b_p} \dots {e_1}_1^{b_1} \hat U
        {e_1}_{a_1}^1 \dots {e_p}_{a_p}^1|0) \: =} \\ && \qqqd
     (0|{e_p}_1^{b_p} \dots {e_1}_1^{b_1} {e_2}_{a_1}^1
        \dots {e_p}_{a_{p-1}}^1 {e_1}_{a_p}^1|0)
     \: = \: \de_{a_p}^{b_1} \de_{a_1}^{b_2} \dots \de_{a_{p-1}}^{b_p}
          \qd, \qd
\eea
such that we can identify $\tau^{(p)}$ with $\hat U$. Let
$F = |\m_1, \dots, \m_s)$. It follows from eq.\ (\ref{tfxxzero}) that
\beq
     \tau^{(p)} F = \prod_{j=1}^s \cot(\m_j) F \qd.
\eeq
Now eq.\ (\ref{xxba}) implies that
\beq \label{proot}
     \left( \prod_{j=1}^s \cot(\m_j) \right)^p = (-1)^{s(s+1)} = 1 \qd,
\eeq
and we have verified that the eigenvalues of $\tau^{(p)}$ are indeed
powers of the $p$th root of unity. Note the the components of $F$
with respect to our basis, which enter the definition (\ref{abastate})
of $|\la_1, \dots, \la_p\>$ can be written as
\beq
     F^{a_1 \dots a_p} = (0|{e_p}_1^{a_p} \dots {e_1}_1^{a_1}
                            B(\m_1) \dots B(\m_s)|0) \qd.
\eeq
        
We can use the symmetries of the monodromy matrices $T(u)$ and
$\CT(u)$ to deduce restrictions on the numbers $p$ and $s$. Eq.\
(\ref{rinv}) implies that the $R$-matrix of the XX chain is invariant
under $\s^z$, $[R,\s^z \otimes I_2 + I_2 \otimes \s^z] = 0$. Let
\beq
     \Si^z = \sum_{j=1}^p \s_j^z \qd.
\eeq
This is twice the operator of the $z$-component of the total spin. We
infer from (\ref{tinv}) that $[T(u),\s^z] = [\Si^z,T(u)]$. $B(u)$,
in particular, commutes with $\Si^z$ as
\beq
     \Si^z B(u) = B(u)(\Si^z - 2) \qd.
\eeq
Thus
\beq \label{szaux}
     \tst{\2} \Si^z |\m_1, \dots, \m_s) =
              \tst{\2}(p - 2s) |\m_1, \dots, \m_s) \qd,
\eeq
$|\m_1, \dots, \m_s)$ is an eigenvector of the $z$-component of the
operator of total spin with eigenvalue $\2(p - 2s)$. This eigenvalue
cannot be smaller than $- \2 p$. Therefore $s \le p$.

Using once more eq.\ (\ref{rinv}) we find that the $R$-matrix of the
Hubbard model in the infinite coupling limit is invariant under the
even generators of gl(1$|$2), $e_1^1$, $e_2^2$, $e_3^3$, $e_3^2$,
$e_2^3$. They span the Lie algebra gl(1)$\oplus$gl(2). The matrix
$e_2^2 + e_3^3$ generates the symmetry operator
\beq
     \hat N = \sum_{j=1}^L \left( {Y_j}_2^2 + {Y_j}_3^3 \right)
            = \sum_{j=1}^L (n_{j \auf} + n_{j \ab}
	                    - 2 n_{j \auf} n_{j \ab}) \qd.
\eeq
This is the particle number operator on our restricted Hilbert space,
where double occupancy of sites is forbidden. Eq.\ (\ref{tinv}) implies
that $[\hat N,\CT(u)] = [\CT(u),e_2^2 + e_3^3]$ and, in particular, that
\beq
     \hat N C_a(u) = C_a(u)(\hat N + 1) \qd.
\eeq
We conclude that
\beq \label{new}
     \hat N |\la_1, \dots, \la_p\> = p |\la_1, \dots, \la_p\> \qd.
\eeq
The state $|\la_1, \dots, \la_p\>$ is constructed by acting on the
vacuum with operators, which leave the Hilbert space with no double
occupancy invariant. Therefore it can contain at most $L$ particles,
and $p \le L$.

Let us perform a naive counting of states. We will start with
the auxiliary spin states $|\m_1, \dots, \m_s)$. Eq.\ (\ref{xxba}) has
$p$ solutions for every $j = 1, \dots s$. The equivalence of XX chain
and a chain of spinless fermions suggests that all $\m_j$ have to be
different. States $|\m_1, \dots, \m_s)$, which merely differ by the
order of the $\m_j$, are identical due to (\ref{bb}). It follows that
we can order the $\m_j$ as $\m_1 < \dots < \m_s$. The number of
solutions of (\ref{xxba}) which satisfy this ordering is $\sum_{s=0}^p
{p \choose s} = 2^p$. Let us assume that the $\la_j$, too, are pairwise
distinct. Then we can order them as $\la_1 < \dots < \la_p$, because
of (\ref{cccom}). The total number of states satisfying the above
restrictions is $\sum_{p=0}^L {L \choose p} 2^p = 3^L$, which is the
dimension of our Hilbert space. This strongly suggests that our
algebraic Bethe ansatz solution provides us with a complete set of
states. An actual proof would require to calculate all scalar products
of the form $\<\la_1, \dots, \la_p|\la_1', \dots, \la_p'\>$, which is
feasible, yet beyond the scope of this article.

We want to point out, that the states $|\la_1, \dots, \la_p\>$ are
eigenstates of the $z$-component of the total spin operator. The proof
goes as follows. The $R$-matrix of the Hubbard model in the infinite
coupling limit is invariant under $\2(e_2^2 - e_3^3)$. Now
\beq
     S^z = \tst{\2} \sum_{j=1}^L (n_{j \auf} - n_{j \ab}) =
           \tst{\2} \sum_{j=1}^L \left( {Y_j}_2^2 - {Y_j}_3^3 \right)
	   \qd,
\eeq
and eq.\ (\ref{tinv}) implies that $[S^z,\CT(u)] = \2 [\CT(u),e_2^2 -
e_3^3]$, or, for the elements $C_a(u)$ of the monodromy matrix,
\beq
     [S^z,C_a(u)] = \tst{\2} \left(\s^z\right)_a^b C_b(u) \qd.
\eeq
It follows that
\bea
     S^z |\la_1, \dots, \la_p\> & = & \tst{\2}
              (\Si^z F)^{a_1 \dots a_p}
	      C_{a_1}(\la_1) \dots C_{a_p}(\la_p)|0\> \nn \\
        & = & \tst{\2} (p - 2s) |\la_1, \dots, \la_p\> \qd. \label{szew}
\eea

Let us compare our results with the recent coordinate Bethe ansatz
solution of the model by Izergin et al.\ \cite{IPA98}. The one-particle
algebraic Bethe ansatz states are easily shown to be of the form
\beq
     C_a(\la) |0\> \propto \sum_{j=1}^L (- \tan(\la))^j
                           c_{j,a}^\dagger |0\> \qd,
\eeq
where $c_{ja} = c_{j \auf}$, if $a = 1$, and $c_{j a} = c_{j \ab}$,
if $a = 2$. From this equation we identify the quasimomenta $k(\la)$
of one-particle states as
\beq
     e^{\i k(\la)} = - \tan(\la) \qd.
\eeq
Let us perform a similar reparametrization for the auxiliary spin
problem,
\beq
     e^{\i q(\m)} = \tan(\m) \qd.
\eeq
Then, using (\ref{proot}), we can rewrite the Bethe ansatz equations
(\ref{baeqn}) and (\ref{xxba}) as
\bea
     e^{\i k_j L} = e^{\i \sum_{k=1}^s q_k} \qd,
                    \qd j = 1, \dots, p \qd, \\
     e^{\i q_k p} = (-1)^{s+1} \qd, \qd k = 1, \dots, s \qd,
\eea
where $k_j = k(\la_j)$, $q_k = q(\m_k)$. These equations agree with the
Bethe ansatz equations obtained in \cite{IPA98}.

Eq.\ (\ref{tau}) implies that the energy $E$ of a state $|\la_1, \dots
\la_p\>$ is
\bea
     E & = & \left. \6_u \ln(\La(u)) \right|_{u=0}
         = \sum_{j=1}^p (\tan(\la_j) + \cot(\la_j)) \nn \\
       & = & - 2 \sum_{j=1}^p \cos(k_j) \qd.
\eea
Eqs.\ (\ref{new}) and (\ref{szew}) show that the states $|\la_1, \dots
\la_p\>$ are also eigenstates of the particle number operator $\hat N$
and the $z$-component of the total spin $S^z$. We may thus add a
chemical potential and a magnetic field to our Hamiltonian
(\ref{hhinf}), $H \rightarrow H_{\m,B}$,
\beq \label{hhinfmub}
     H_{\m, B} = \sum_{j=1}^L \{ - \D (c_{j \s}^\dagger c_{j+1, \s} +
                         c_{j+1, \s}^\dagger c_{j \s}) \D
			 \, - \, \m (n_{j \auf} + n_{j \ab}) \,
			 + \, B (n_{j \auf} - n_{j \ab}) \} \, .
\eeq
Then $H_{\m, B} |\la_1, \dots, \la_p\> = E_{\m, B} |\la_1, \dots,
\la_p\>$, where
\beq
     E_{\m, B} = - 2 \sum_{j=1}^p \cos(k_j) - \m p + B(p -2s) \qd.
\eeq
This means that the algebraic Bethe ansatz states $|\la_1, \dots,
\la_p \>$ can be used to calculate form factors and correlation
functions in the grand canonical ensemble (cf \cite{IPA98}).
\subsection*{Yangian symmetry of the Hubbard model in the infinite
coupling limit}
The Hubbard Hamiltonian on the infinite interval is invariant under
the direct sum of two su(2) Yangian quantum groups
\cite{UgKo94,MuGo97a}. We would like to point out here, that one of
these Yangian symmetries survives the infinite coupling limit.
First of all, let us recall \cite{Drinfeld85} that the su(2) Yangian
is a Hopf algebra which is spanned by two triples of generators
$I^a$, $J^a$, $a = x, y, z$, satisfying the relations
\bea
     [I^a,I^b] & = & c_{abc} I^c \qd, \\[0mm]
     [I^a,J^b] & = & c_{abc} J^c \qd, \\[0mm] \nn
     [[J^a,J^b],[I^c,J^d]] & + & [[J^c,J^d],[I^a,J^b]] = \\
     \label{serre} &&
     -4 (a_{abefgh} c_{cde} + a_{cdefgh} c_{abe}) \{I^f,I^g,J^h\} \qd.
\eea
Here $c_{abc} = \i \e^{abc}$ is the antisymmetric tensor of structure
constants of su(2), and $a_{abcdef} = c_{adg} c_{beh} c_{cfi} c_{ghi}$.
The bracket $\{\:\}$ in (\ref{serre}) denotes the symmetrized product
\beq
     \{x_1,x_2,x_3\} = \dst{\frac{1}{6}}
                       \sum_{i \ne j \ne k \ne i} x_i x_j x_k \qd.
\eeq
Being a Hopf algebra Y(su(2)) carries an outer structure
(co-multiplication, antipode, co-unit), which guarantees that 
Y(su(2)) has a rich representation theory \cite{ChPr90}.

As a corollary of a theorem proven in \cite{GoIn96} it follows that
\bea
     I^a & = & \sum_j S_j^a \qd, \\
     J^a & = & \sum_{j \ne k} \sign(j - k) \e^{abc} S_j^b S_k^c
\eea
form a representation of Y(su(2)). The $S_j^a$ in this equation are
spin density operators in fermionic representation (\ref{defs}).
$I^a$ obviously commutes with the Hubbard Hamiltonian in the infinite
coupling limit (\ref{hhinf}). We will give a simple proof
\cite{Inozemtsev95} that $J^a$, too, commutes with the Hamiltonian
(\ref{hhinf}), if we replace the summation in (\ref{hhinf}) by a
summation over all integers. Let $S_{jk}^a = \2 \s_{\a \be}^a
c_{j \a}^\dagger c_{k \be}$, and define
\beq
     K^a = 2 \i \sum_j (S_{j, j+1}^a - S_{j+1, j}^a) \qd. 
\eeq
It was shown in \cite{UgKo94,GoIn96} that the Hubbard Hamiltonian
$H_H$,
\bea
     H_H & = & T + UD \qd, \\
     T & = & - \sum_j (c_{j \s}^\dagger c_{j+1 \s}
             + c_{j+1 \s}^\dagger c_{j \s}) \qd, \\
     D & = & \sum_j n_{j \auf} n_{j \ab} \qd,
\eea
commutes with
\beq \label{hyang}
     J_U^a = J^a + U^{-1} K^a
\eeq
and that $I^a$ and $J_U^a$ form a representation of the Y(su(2))
Yangian. Since $J_U^a$ commutes with $H_H$ for all real $U$, we obtain
the identity $[T,J^a] = [K^a,D]$. Now $D \D = \D D = 0$, and
$[\D,J^a] = 0$, since $[\D,S_j^a] = 0$. It follows that
\beq
     [H,J^a] = [\D T \D, J^a] = \D [T,J^a] \D = \D [K^a,D] \D = 0 \qd.
\eeq

Let us remark that a more systematic way to explore the Yangian
symmetry of the model would be to apply the approach developed in
\cite{MuWa96,MuGo97a,MuGo98}, where the Yang-Baxter algebra was
investigated in the thermodynamic limit. This approach would also
provide us with the action of the Yangian on eigenstates.
\subsection*{Conclusions}
We would like to stress that our approach is widely applicable. Given
a solution $R(u,v)$ of the Yang-Baxter equation (\ref{ybe}), it
consists of the following steps.
\begin{enumerate}
\item
Use eq.\ (\ref{comp}) to choose a grading which is compatible with the
$R$-matrix. In general, this choice is not unique. A given $R$-matrix
may be compatible with different gradings.
\item
For a given grading choose a fermionic representation. Start with a
tensor product representation  (\ref{multipro}) of projection operators
${Y_j}_\a^\be$ of sufficiently high dimension and adjust it to the
grading by deleting rows and columns of ${Y_j}_\a^\be$. Again the
choice is not unique.
\item
Write down the Hamiltonian (\ref{hdens}) and replace the graded spin
operators ${e_j}_\a^\be$ by fermionic projection operators
${Y_j}_\a^\be$. This Hamiltonian is integrable by construction. There
may be different methods to diagonalize it.
\item
The Hamiltonian can be diagonalized by (nested) coordinate Bethe ansatz.
\item
It may be possible to diagonalize the Hamiltonian algebraically. In this
case the $L$-matrix is obtained from eq.\ (\ref{defgl}), and the graded
Yang-Baxter algebra is given by (\ref{gyba}). Note however, that there
is no general recipe for an algebraic Bethe ansatz. Depending on the
structure of the $R$-matrix and of the monodromy matrix an algebraic
Bethe ansatz may be a difficult task (cf e.g.\
\cite{Tarasov88,RaMa97}), if it is possible at all.
\end{enumerate}
Our approach circumvents certain difficulties connected with the
Jordan-Wigner transformation, which, in certain cases, may be
alternatively used to connect a spin representation of an integrable
system with a fermionic representation. In our approach there is no
twisting of the boundary conditions, and symmetries are directly
obtainable via eq.\ (\ref{rinv}). Our approach works for higher
su($N$) spins where no Jordan-Wigner transformation is known.

The main example to illustrate our ideas was the Hubbard model in the
infinite coupling limit. An interesting lesson to learn from this
example is that the Hubbard model in the infinite coupling limit is
equivalent to a su(3)-spin generalization of the XX spin chain.
\subsection*{Acknowledgements}
F.\ G.\ acknowledges partial financial support by the Deutsche
Forschungsgemeinschaft under grant number Go 825/2-1.


\begin{thebibliography}{10}

\bibitem{KuSk82}
P.~P. Kulish and E.~K. Sklyanin, J. Soviet Math. {\bf 19}, 1596 (1982).

\bibitem{Kulish85}
P.~P. Kulish, J. Soviet Math. {\bf 35}, 2648 (1985).

\bibitem{BGZD94}
A.~J. Bracken, M.~D. Gould, Y.-Z. Zhang, and G.~W. Delius,
J. Phys. A {\bf 27}, 6551 (1994).

\bibitem{DGLZ95}
G.~W. Delius, M.~D. Gould, J.~R. Links, and Y.-Z. Zhang,
Int. J. Mod. Phys. A {\bf 10}, 3259 (1995).

\bibitem{Maassarani95}
Z.~Maassarani, J. Phys. A {\bf 28}, 1305 (1995).

\bibitem{Shastry88b}
B.~S. Shastry, J. Stat. Phys. {\bf 50}, 57 (1988).

\bibitem{OWA87}
E.~Olmedilla, M.~Wadati, and Y.~Akutsu, J. Phys. Soc. Jpn. {\bf 56},
2298 (1987).

\bibitem{MaMa97a}
Z.~Maassarani and P.~Mathieu, Nucl. Phys. B {\bf 517}, 395 (1998).

\bibitem{Luescher76}
M.~L\"{u}scher, Nucl. Phys. B {\bf 117}, 475 (1976).

\bibitem{GoMu97b}
F.~G\"ohmann and S.~Murakami, J. Phys. A {\bf 30}, 5269 (1997).

\bibitem{SUW98}
M.~Shiroishi, H.~Ujino, and M.~Wadati, J. Phys. A {\bf 31}, 2341 (1998).

\bibitem{Lai74}
C.~K. Lai, J. Math. Phys. {\bf 15}, 1675 (1974).

\bibitem{Sutherland75}
B.~Sutherland, Phys. Rev. B {\bf 12}, 3795 (1975).

\bibitem{Schlottmann87}
P.~Schlottmann, Phys. Rev. B {\bf 36}, 5177 (1987).

\bibitem{BaBl90}
P.~A. Bares and G.Blatter, Phys. Rev. Lett. {\bf 64}, 2567 (1990).

\bibitem{Sarkar91}
S.~Sarkar, J. Phys. A {\bf 24}, 1137 (1991).

\bibitem{EsKo92}
F.~H.~L. E{\ss}ler and V.~E. Korepin, Phys. Rev. B {\bf 46}, 9147
(1992).

\bibitem{FoKa93}
A.~Foerster and M.~Karowski, Nucl. Phys. B {\bf 396}, 611 (1993).

\bibitem{EKS92}
F.~H.~L. E{\ss}ler, V.~E. Korepin, and K.~Schoutens,
Phys. Rev. Lett. {\bf 68}, 2960 (1992).

\bibitem{EKS94}
F.~H.~L. E{\ss}ler, V.~E. Korepin, and K.~Schoutens,
Int. J. Mod. Phys. B {\bf 8}, 3205 (1994).

\bibitem{KBIBo}
V.~E. Korepin, N.~M. Bogoliubov, and A.~G. Izergin,
{\em Quantum Inverse Scattering Method and Correlation Functions},
Cambridge University Press, (1993).

\bibitem{IPA98}
A.~G. Izergin, A.~G. Pronko, and N.~I. Abarenkova.
\newblock Temperature correlators in the one-dimensional Hubbard
model in the strong coupling limit.
\newblock PDMI preprint 5/1998, hep-th/9801167,  (1998).

\bibitem{UgKo94}
D.~B. Uglov and V.~E. Korepin, Phys. Lett. A {\bf 190}, 238 (1994).

\bibitem{MuGo97a}
S.~Murakami and F.~G\"ohmann, Phys. Lett. A {\bf 227}, 216 (1997).

\bibitem{Drinfeld85}
V.~G. Drinfel'd, Soviet Math. Dokl. {\bf 32}, 254 (1985).

\bibitem{ChPr90}
V.~Chari and A.~Pressley, L'Enseignement Math\'ematique {\bf 36},
267 (1990).

\bibitem{GoIn96}
F.~G\"ohmann and V.~Inozemtsev, Phys. Lett. A {\bf 214}, 161 (1996).

\bibitem{Inozemtsev95}
V. Inozemtsev,
\newblock private communication,  (1995).

\bibitem{MuWa96}
S.~Murakami and M.~Wadati, J. Phys. A {\bf 29}, 7903 (1996).

\bibitem{MuGo98}
S.~Murakami and F.~G\"ohmann, Nucl. Phys. B {\bf 512}, 637 (1998).

\bibitem{Tarasov88}
V.~O. Tarasov, Theor. Math. Phys. {\bf 76}, 793 (1988).

\bibitem{RaMa97}
P.~B. Ramos and M.~J. Martins, J. Phys. A {\bf 30}, L195 (1997).

\end{thebibliography}

\end{document}